\documentclass[a4paper]{jpconf}
\usepackage{graphicx}
\usepackage[english]{babel}   
\usepackage[usenames]{color}   
\usepackage{colortbl} 
\usepackage[normalem]{ulem}  %

\begin{document}
\title{Formation of spiral dwarf galaxies: observational data and results of numerical simulation
\footnote{\color{blue}\underline{\bf Published}: Khrapov S.S., Khoperskov A.V., Zaitseva N.A., Zasov A.V., Titov A.V., Formation of spiral dwarf galaxies: observational data and results of numerical simulation // St. Petersburg State Polytechnical University Journal. Physics and Mathematics, 2023, v.16 (1.2), pp.395--402. DOI: https://doi.org/10.18721/JPM.161.260  https://physmath.spbstu.ru/en/article/2023.64.60/}}

\author{Sergey Khrapov$^1$, Alexander Khoperskov$^1$, Natalia Zaitseva$^3$, Anatoly~Zasov$^{2,3}$, Alexander Titov$^1$} 
\address{$^1$ Institute of Mathematics and Information Technology, Volgograd State University, Volgograd, Russia \\
         $^2$ Lomonosov Moscow State University, Faculty of Physics, Moscow, Russia \\
         $^3$ Lomonosov Moscow State University, Sternberg Astronomical Institute, Moscow, Russia
        }

\ead{khrapov@volsu.ru, khoperskov@volsu.ru}

\begin{abstract}
Recent studies show the possibility of the formation of fairly regular and global spiral patterns in dwarf galaxies (dS type).
Our sample of observed dwarf objects of this class also includes galaxies with a central stellar bar.
The analysis of the observational data provides a small rotation velocity and a small disk component mass for dS galaxies, which is in poor agreement with the spiral structure generation mechanism in isolated dwarfs due to the development of disk gravitational instability.
Numerical simulation of the stellar–gaseous disks self-consistent dynamics imposes restrictions on the stellar disk thickness and the maximum gas rotation velocity, at which the gravitational mechanism of spiral formation can still be effective. 
\end{abstract}

\section{Introduction}
Dwarf galaxies are small in size and mass compared to classical spiral galaxies (S or SB types) and are usually considered structureless, irregular objects (Irr).
Some late-type dwarfs (Sd~--~Sm types) have a rotating stellar disk without any regular and developed spiral structure.
Such galaxies exhibit flocculent type of spirals, which are discontinuous and consist of short regions~\cite{Mondal-etal-2021}.
Gravitational instability in large massive disks is able to provide a regular spiral pattern covering the entire disk, so Grand Design spiral structure is common for Sa -- Sc galaxy types \cite{Kormendy-2011rev, khrapov-etal-2021galaxies, Griv-etal-2017, Dobbs-Baba-2014spiral, Buta-2013rev}. 

Only a small part of dwarf galaxies shows a global, relatively regular spiral pattern in their disk, and such objects belong to the fairly rare dS type \cite{Zasov-etal-2021dwarf, Magana-Serrano-etal-2020dwarf-dS, Edmunds-Roy-1993-dS}. The observations comparative analysis of normal S- and dS-galaxies represents that such dwarfs are more than just smaller copies of large objects, since their spectral characteristics are similar to Irr galaxies \cite{Hidalgo-Gamez-2004-dS}.
The small size and mass of dwarf galaxies appear to be a theoretical problem for the formation of extended spirals due to gravitational instability \cite{Dobbs-Baba-2014spiral, Buta-2013rev}.

Here we consider the observed properties of the sample of dS galaxies compared to dwarf galaxies without a regular spiral structure \cite{Zasov-etal-2021dwarf}.
The numerical simulation of the dynamics of dwarfs stellar disks with a rich gaseous component makes it possible to determine the conditions under which gravitational instability can generate sufficiently extended spiral patterns in dS galaxies, which morphology is similar to Grand Design galaxies.

\begin{figure}[h]
\includegraphics[width=0.99\hsize]{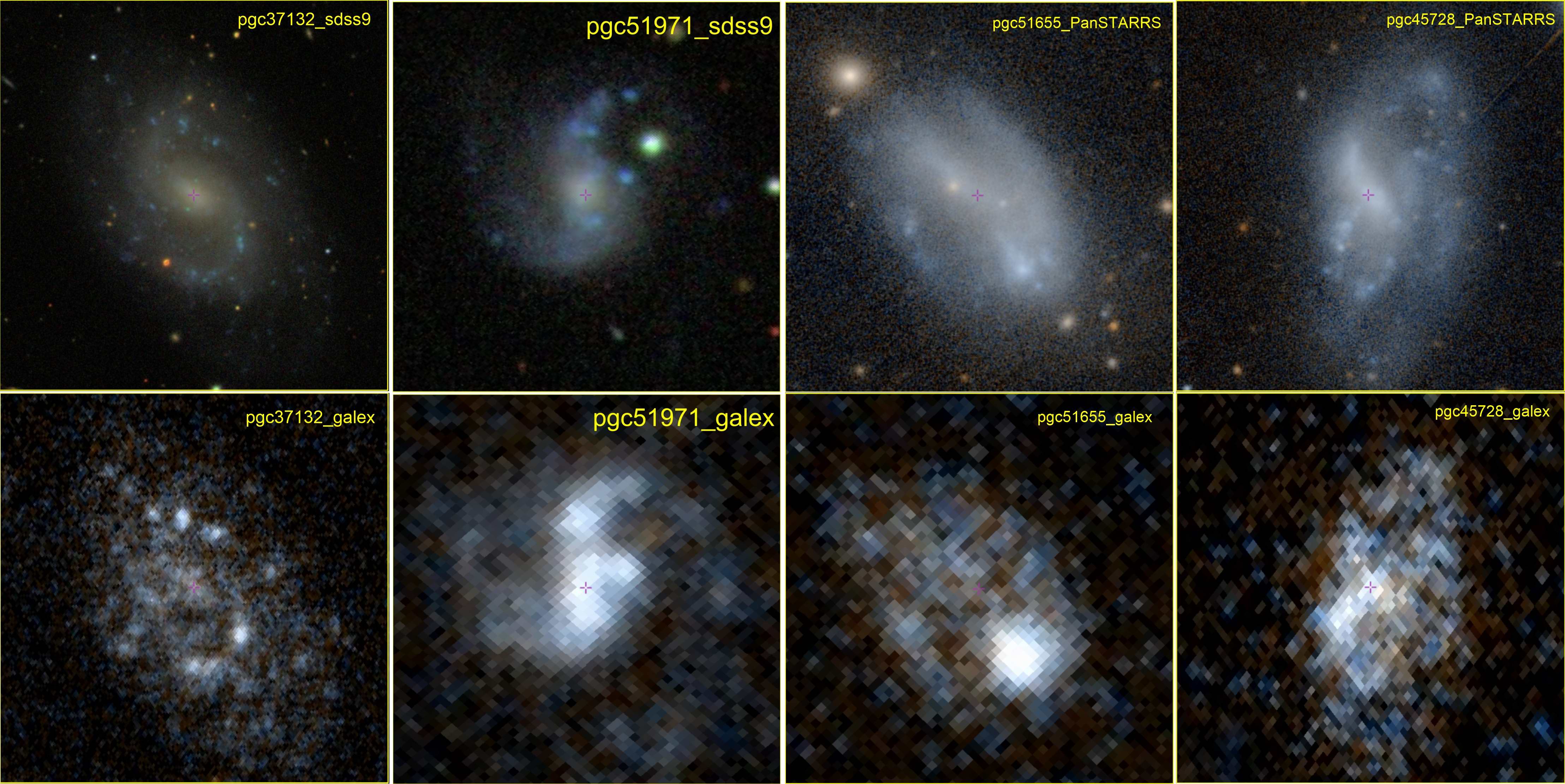}
\caption{\label{fig:ImageObservation1}
Dwarf galaxies with a spiral pattern from our sample.}
\end{figure}

\section{Sample of dS-galaxies and its properties}

\begin{figure}[h]
\includegraphics[width=0.99\hsize]{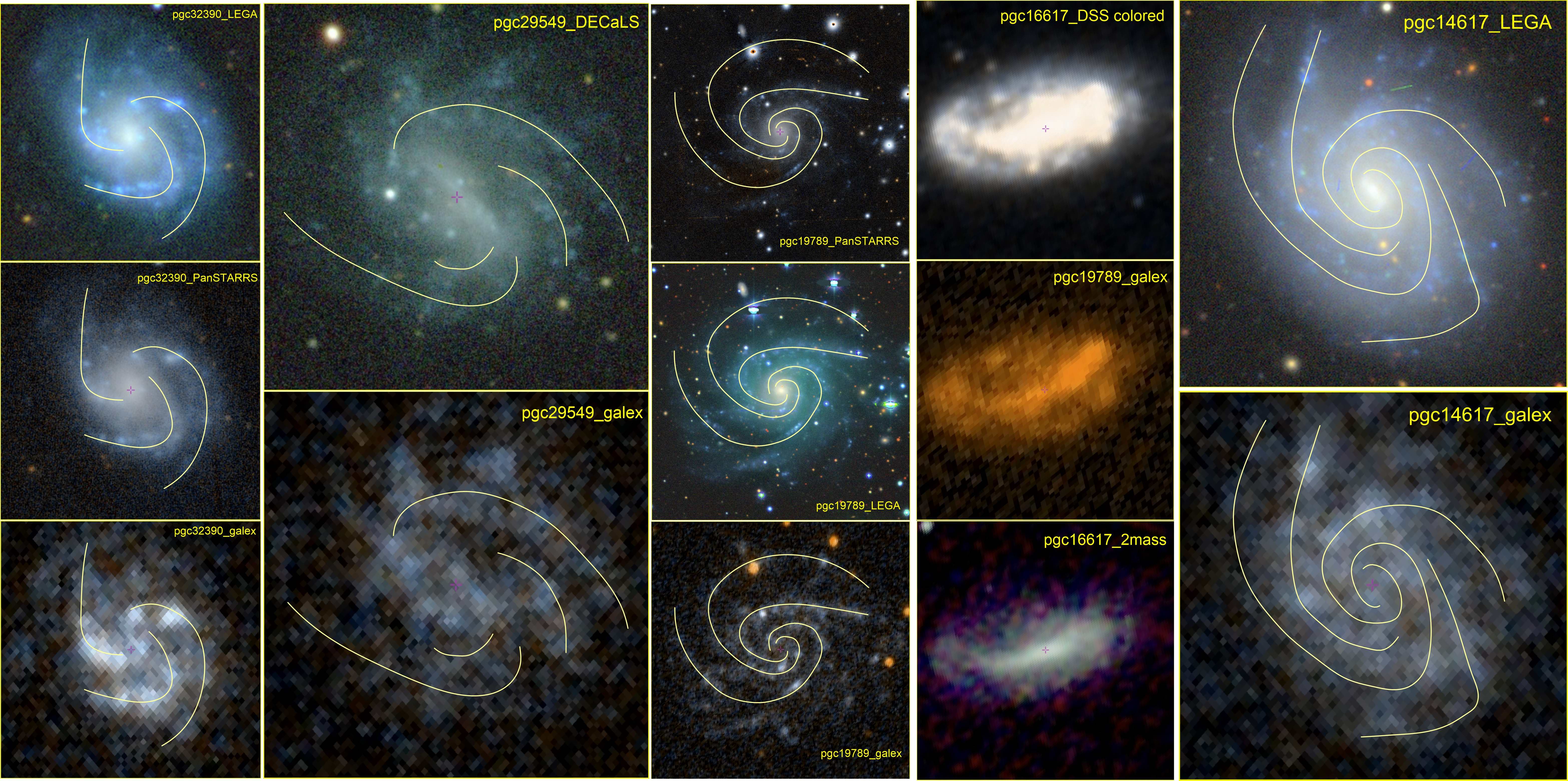}
\caption{\label{fig:ImageObservation2}
Dwarf galaxies with a spiral pattern from our sample (continued).}
\end{figure}

Our sample is limited to objects (usually type Sc -- Irr) with the absolute magnitude $M_B > -18^{m}$, the optical diameter $D_{25} < 12$\,kpc, $m_B < 15^m$, the inclination angle $i< 75^\circ$ \cite{Zasov-etal-2021dwarf}. 
It is also important to have images in different spectral ranges for deeper study (Figures \ref{fig:ImageObservation1}, \ref{fig:ImageObservation2}).
The logarithm of isolation index $\log(ii)$ characterizes the degree of environmental influence \cite{Makarov-Karachentsev-2011interactions} and we do not consider both Virgo, UMa, Fornax clusters and peculiar galaxies with signs of strong interaction.

Our sample of spiral dwarf galaxies includes 43 objects, which are compared with the sample of dwarfs without spirals (119 objects of Sm and Irr types, see detailed description in \cite{Zasov-etal-2021dwarf}).
Figure~\ref{fig:ImageObservation1} shows dwarf galaxies with bars and two distinct arms.
The images of SDSS\,9 (the Sloan Digital Sky Survey), DECaLS (the Dark Energy Camera Legacy Survey), DSS (the Digitized Sky Survey), PanSTARRS (the Panoramic Survey Telescope and Rapid Response System), LEGA (the DESI Legacy Imaging Surveys) \cite{Dey-etal-2019LEGA} characterize the distributions of the stellar components.
The bottom rows in the fig.~\ref{fig:ImageObservation1},~\ref{fig:ImageObservation2} show the distributions of gas and young stars according to GALEX data (the Galaxy Evolution Explorer).
Figure \ref{fig:ImageObservation2} demonstrates galaxies with a more complex spiral structure, where the yellow lines highlight the positions of the spiral arms. Moreover, the spirals in the stellar and gaseous components are in good agreement with each other. Three-arm patterns indicate a rather massive dark halo.

\begin{figure}[h]
\includegraphics[width=0.995\hsize]{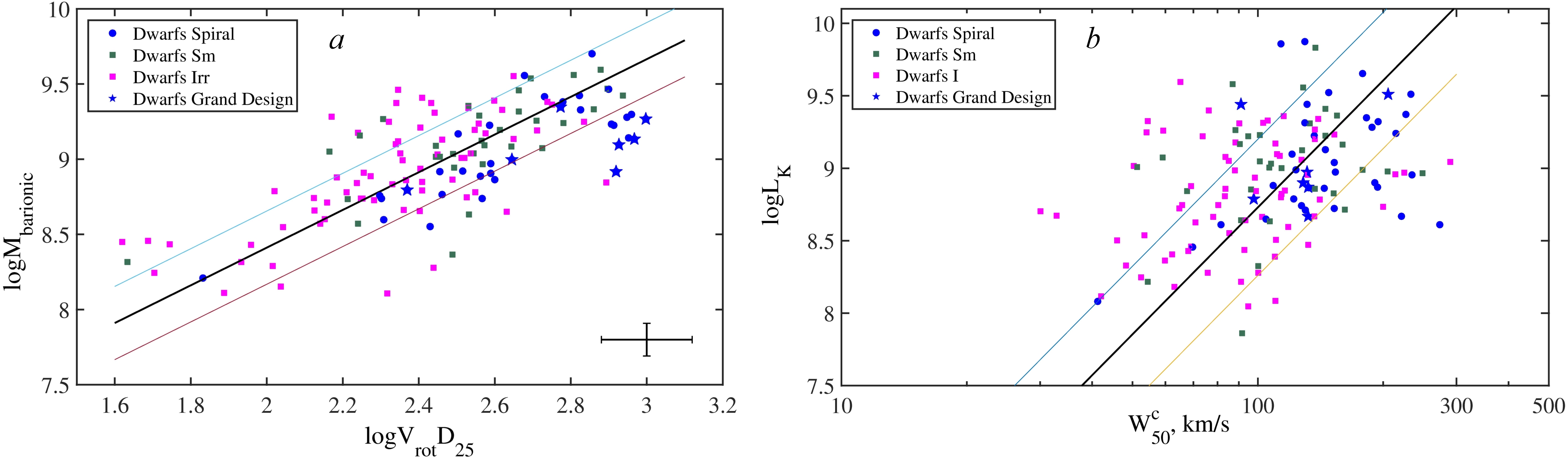}
\caption{\label{fig:ObservationStatist}
Positions of various dwarf galaxies: blue icons are spirals, green squares are Sm-type, pink squares are irregular objects.
\textit{a}) In-plane distribution of the ``baryon mass -- momentum'' parameters compared to the regression for isolated galaxies of the AMIGA sample \cite{Lisenfeld_etal_07} (black line), colored lines show deviation 1$\sigma$, $ M_{barionic} = \Upsilon^*_KL_K+\eta M_{HI}$, where $\Upsilon^*_K=0.6$ \cite{McGaugh_Schombert} and $\eta=1.33$.
\textit{b}) Tully-Fisher relation for the K-band compared to those obtained by Karachentsev et al. \cite{Karachentsev_etal_17} for the sample of Local Volume dwarf galaxies (black line, colored lines show 1$\sigma$~deviation). 
 }
\end{figure}

Each galaxy in our two samples is characterized by the systemic velocity $V_{sys}$, the diameter $D_{25}$, the maximum rotation velocity $V_{rot}$, the HI mass $M_{HI}$, the estimate of the total gravitating mass $M_{dyn}$, the luminosities $L_K$ according to the K-magnitudes of 2MASS catalog \cite{Huchra-etal-2012-2mass}. Statistical analysis gives the following conclusions (See also \cite{Zasov-etal-2021dwarf}).

\noindent --- Dwarf galaxies with developed spiral structure are the most massive objects in the sample.

\noindent --- The distributions of dS galaxies and objects without spirals indicate the absence of very significant differences for various pairs of parameters, for example, $L_K - M_{dyn}$, $L_K - M_{HI}$, $D_{25} - M_{HI}$, $M_{dyn} - M_{HI}$, $V_{rot} - M_{HI}$ and others (Figure \ref{fig:ObservationStatist}).

\noindent --- The HI mass in dS galaxies is, on average, about two times less than in dwarf non-spiral galaxies, although the dynamical and photometric parameters are close for both samples.

\noindent --- The central stellar bar is found both in dS-type objects and in non-spiral galaxies.

\noindent --- Tidal influence is apparently not an essential factor of the  formation in considered galaxies.

\noindent --- The proportion of baryonic matter in spiral dwarfs is, on average, lower than in objects with irregular structure and in giant spiral galaxies, which may indicate the influence of the dark halo on the formation of the spiral patterns in dS-dwarfs.

\noindent --- Remote dwarf galaxies follow the same Tully-Fisher relation as Local Volume dwarfs, but their physical parameters are determined with a larger uncertainties due to the low brightness of these objects, which significantly increases the points scatter on the diagram.

\begin{figure}[h] \begin{center}
\includegraphics[width=0.495\hsize]{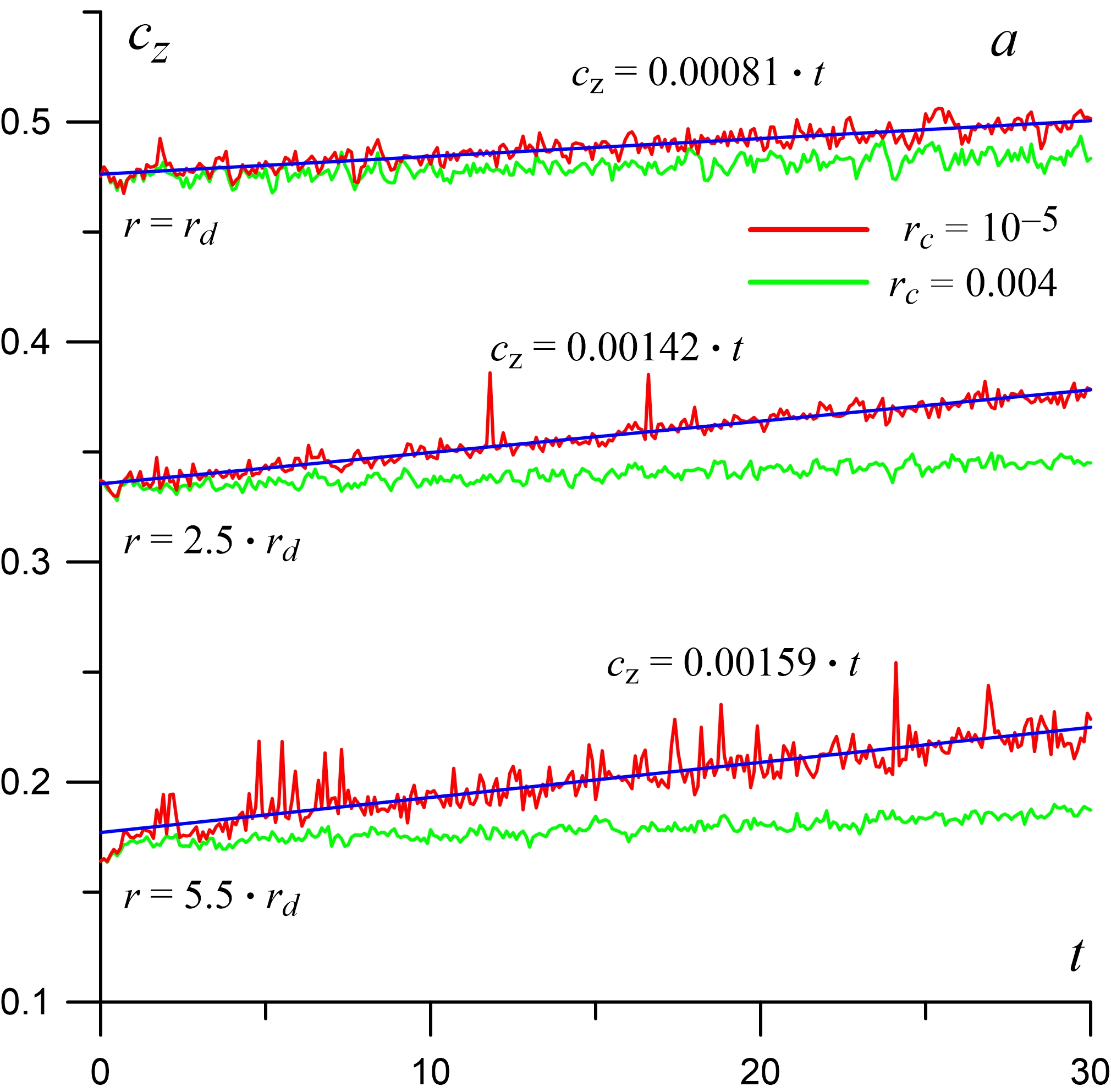} \includegraphics[width=0.495\hsize]{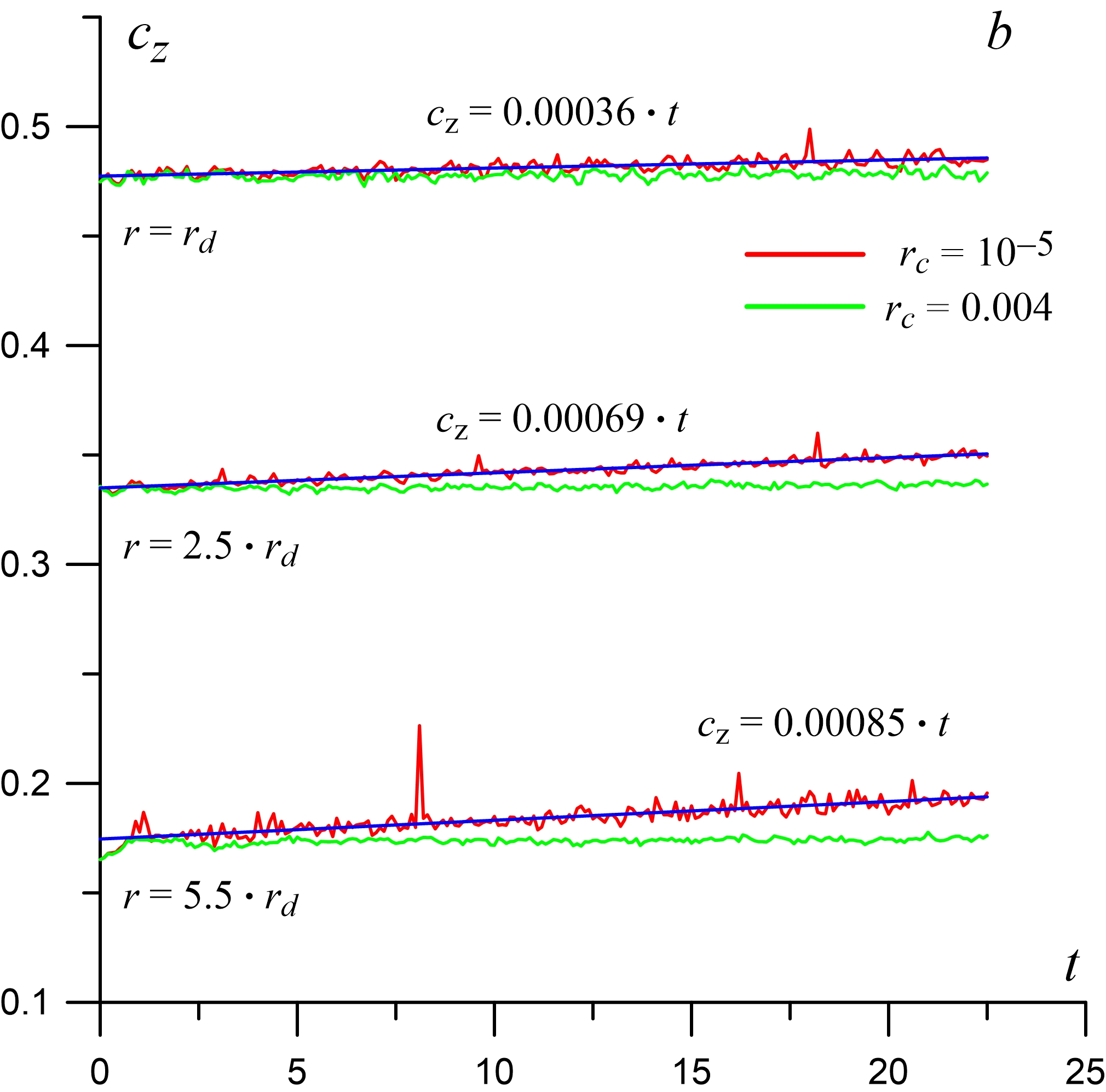} \end{center}
\caption{\label{fig:Relaxation}
The dispersion dynamics of stellar vertical velocities in disk models for different gravitational potential cutoff radii ($r_c$), the number of particles in the disk ($N^{(*)}$) and the live halo ($N^{(h)}$).
\textit{a}) $N^{(*)} = N^{(h)} =  2^{18}$, $r=10^{-5}$ (red lines), $r=4\cdot 10^{-3}$ (green lines), for different radii ($r_d$~is the radial exponential scale of stellar disk);   
 \textit{b}) $N^{(*)} = N^{(h)} =  2^{20}$, $r=10^{-5}$ (red lines), $r=4\cdot 10^{-3}$ (green lines), for different radii.    
 }
\end{figure}

\section{Numerical modeling of the galactic disk}

The numerical models are based on the self-consistent dynamics of the N-body gravitating system for the stellar disk and the gaseous component, which is described by the hydrodynamic equations \cite{khrapov-etal-2021galaxies, Zasov-etal-2021dwarf}.
We used direct method to calculate self-gravity forces (each particle interacts with each other), which is the most accurate modeling approach \cite{Khrapov-2018chel}. The GPUs application for parallel code makes it possible to perform fairly fast numerical experiments with $N = 2^{20}-2^{23}$.
The numerical model should ensure the collisionlessness of the stellar component \cite{Zasov-etal-2021dwarf, Smirnov-Sotnikova-2018relax}, which is achieved by cutting off the Newtonian potential at small radii $r_c$.
Surface density of the stellar exponential disk $\sigma(r) = \sigma_0 \, \exp(-r/r_d)$ characterized by the radial scale $r_d$. We use dimensionless units of length $r=1 \rightarrow 4$\,kpc, velocity $V=1 \rightarrow 47$\,km\,sec$^{-1}$ and time $t=1 \rightarrow 80$\,Myr.

Figure \ref{fig:Relaxation} demonstrates the presence or absence of the stellar disk heating for the corresponding values of $r_c$ and $N = N^{(*)} + N^{(h)}$ (where $N^{(*)} $ is the number of particles in the stellar disk, $N^{(h)}$~is the number of particles that form dark live halo). 

 We see a noticeable linear increase in vertical velocity dispersion at a very small cutoff radii due the absence of collisionlessness in such a model. The value $r_c = 0.004$ ensures the almost stationary behavior of the velocity dispersions (green lines). 
The heating is stronger for a smaller values of the number of particles $N^{(*)}$ and $N^{(h)}$ (comparison of Figure \ref{fig:Relaxation}a and Figure \ref{fig:Relaxation}b).

\begin{figure}[h] \begin{center}
\includegraphics[width=0.995\hsize]{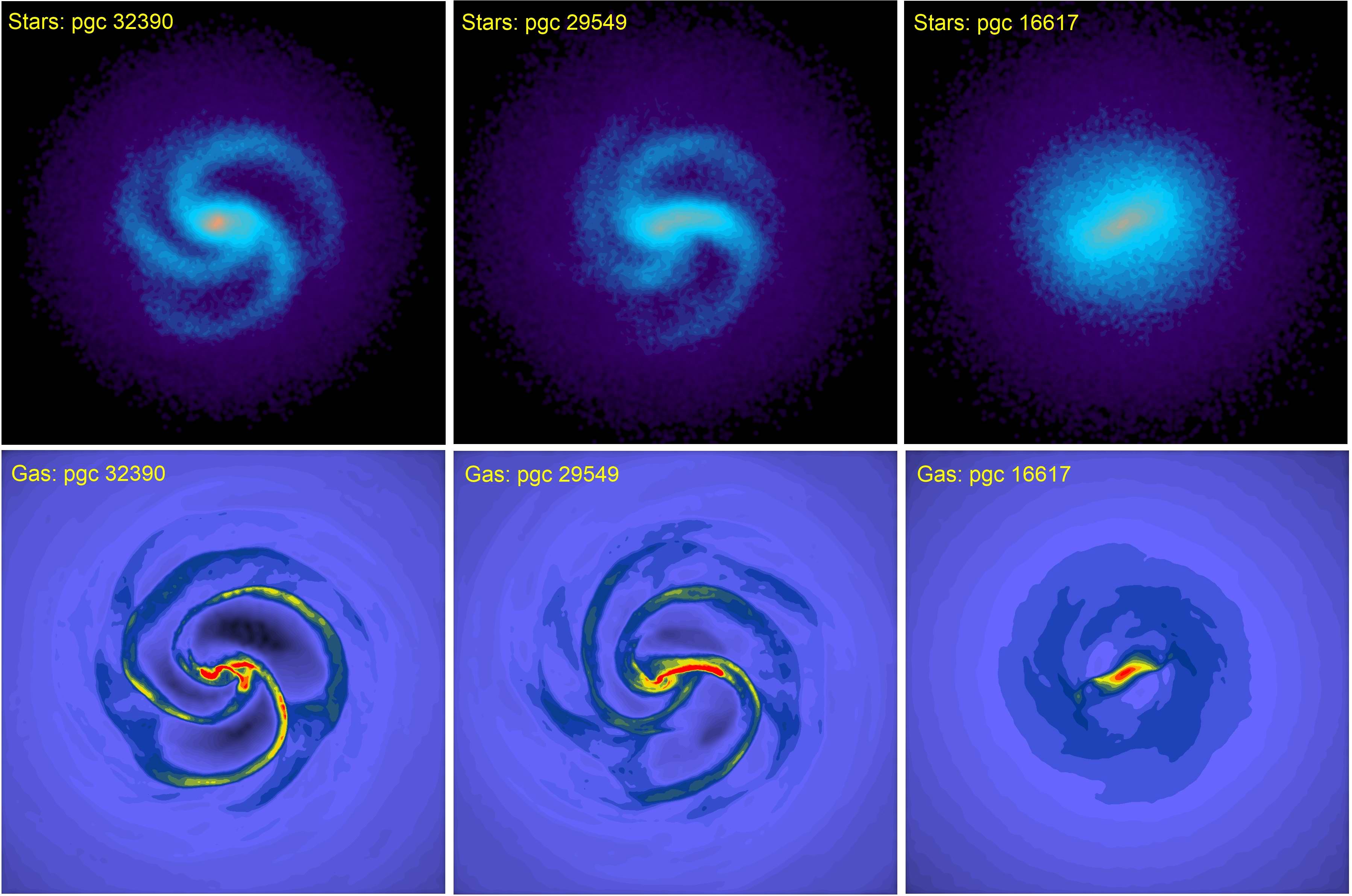} \end{center}
\caption{\label{fig:NumericalModels}
Examples of spiral patterns in numerical models of galaxies PGC~32390, PGC~29549, PGC~16617. Top row: stellar disk, bottom row: gaseous disk.}
\end{figure}

The figure \ref{fig:NumericalModels} shows various spiral structures in dwarf galaxies models.
We obtain different stellar and gaseous disks morphology in numerical models by varying the relative masses of stars, gas, and dark halo, as well as the radial and vertical scales that determine the distributions of subsystem parameters. The model structures in gaseous and stellar disks are close to the observed patterns of the corresponding galaxies (See Figures~\ref{fig:ImageObservation1}, \ref{fig:ImageObservation2}). Other calculation examples are given in \cite{Zasov-etal-2021dwarf}.

\section{Conclusion} 

\begin{figure}[h] \begin{center}
\includegraphics[width=0.9\hsize]{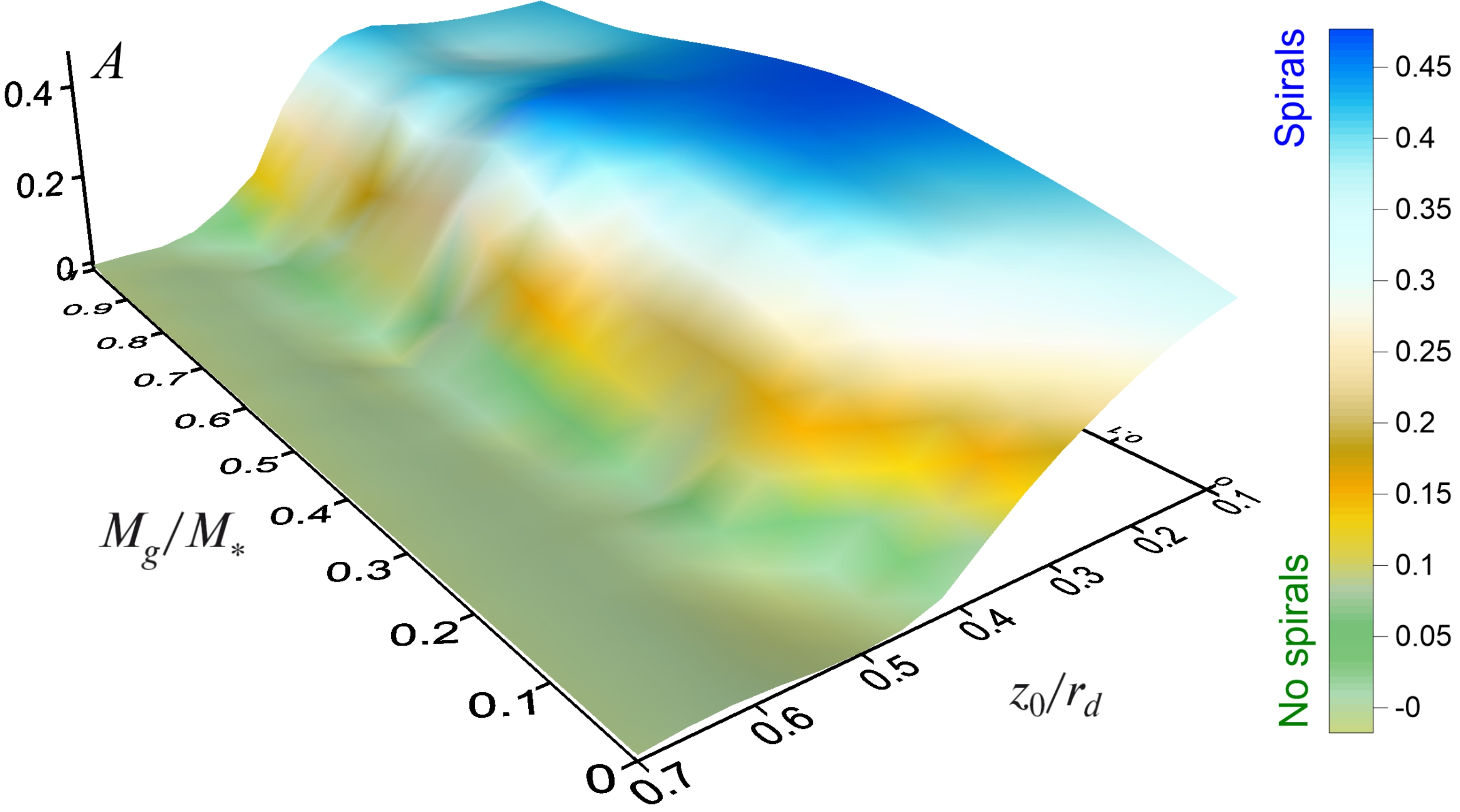} \end{center}
\caption{\label{fig:FourierHarmonicsAmplitude}
Distribution of the Fourier harmonics maximum amplitude $A$ on the plane of dimensionless parameters (where  $M_g/M_*$ is the relative gas mass, $z_0$ is the vertical disk scale, $r_0$ is the radial disk scale).
 }
\end{figure}

Photometric and kinematic observational data do not allow us to confidently identify the factors that ensure the formation of a global spiral patterns in dwarf galaxies, which are a rather rare phenomenon.
There is only some gas deficit in dS galaxies compared to dIrr objects.

We have studied the possibility of the global spiral structure formation in numerical models of isolated dwarf galaxies due to the development of gravitational instability in the stellar disk rich in gas.
The presence of a spiral pattern in dwarf models imposes some restrictions on the disks thickness, the radial velocity dispersion profiles in stellar disk, the sound speed in gas and the gas density. 
The results of numerical simulations show that the maximum gas rotation velocity must be higher than 60 km\,sec$^{-1}$ in order to excite spiral waves with significant amplitude.
Thicker stellar disk requires more gas to form the spiral pattern.

\subsection*{Acknowledgments}
This research has made use of ``Aladin sky atlas'' developed at CDS, Strasbourg Observatory, France. The work was supported by the Ministry of Science and Higher Education of the Russian Federation (state assignment, project No. 0633-2020-0003, implementation of all numerical simulations) and by Russian Foundation for Basic Research (project 20-02-00080~A, observational data analysis).

\end{document}